\documentclass[artile]{aa}  

\usepackage{nameref}

\usepackage{natbib}
\usepackage{booktabs}
\usepackage[para]{threeparttable} 
\usepackage{threeparttable} 
\usepackage{multirow} 

\bibpunct{(}{)}{;}{a}{}{,}

\usepackage{amsmath,amssymb,amsfonts}
\usepackage{graphicx}
\usepackage{txfonts}
\usepackage{xcolor}
\usepackage{stfloats}
\usepackage{placeins}

\begin{document}

\title{Detection of OH and Fe in the dayside atmosphere of the hottest ultra-hot Jupiter KELT-9b with SPIRou}

\author
{Yuanheng Yang\inst{1,2}
\and Guo Chen\inst{1,3}
\and Hui Zhang\inst{4}
\and Fei Yan\inst{5}
}

\institute{
CAS Key Laboratory of Planetary Sciences, Purple Mountain Observatory, Chinese Academy of Sciences, Nanjing 210023, China\\
\email{yhyang@pmo.ac.cn; guochen@pmo.ac.cn}
\and School of Astronomy and Space Science, University of Science and Technology of China, Hefei 230026, China
\and CAS Center for Excellence in Comparative Planetology, Hefei 230026, China
\and Shanghai Astronomical Observatory, Chinese Academy of Sciences, Shanghai 200030, China
\and Department of Astronomy, University of Science and Technology of China, Hefei 230026, China
}

\date{Received xxx; accepted xxx}

\titlerunning{Detection of OH and Fe in the dayside atmosphere of the hottest ultra-hot Jupiter KELT-9b with SPIRou}

\authorrunning{Yang et al.}

\abstract 
{
Simultaneous abundance measurements of volatile and refractory elements are crucial to unravelling the formation and migration history of ultra-hot Jupiters (UHJs). High-resolution infrared emission spectroscopy has recently been employed extensively to investigate the atmospheric components of UHJs, including both molecules and atoms. For the hottest known planet, KELT-9b, whose dayside atmosphere is almost completely thermally dissociated and ionized, no molecular components have been conclusively detected. Here we present the first detection of the OH molecule in the dayside atmosphere of KELT-9b, based on two thermal emission observations conducted with the SPIRou spectrograph, and confirm the presence of Fe in the dayside hemisphere. We performed a self‐consistent retrieval under the assumption of chemical equilibrium, constraining elemental abundances and atmospheric metallicity ([M/H]). We confirm the presence of a significant thermal inversion layer on the dayside. By retrieval, no significant net Doppler shift signals are identified, and the retrieved equatorial rotation speed agrees with the tidally locked rotation speed. The retrieved oxygen abundance is solar to supersolar ($0.61_{-0.58}^{+1.19} \mathrm{dex}$). The retrieval suggests a subsolar to solar [C/O] ($-0.75_{-0.82}^{+0.64} \mathrm{dex}$) and a subsolar to solar atmospheric metallicity. The low metallicity may point to a locally well-mixed envelope and interior. The constraints remain broad, and the data are still statistically consistent with supersolar C/O and subsolar oxygen abundances. Taken together, the [C/O] and [O/H] results are compatible with formation beyond the water snowline followed by inward migration, but the present data do not conclusively rule out other scenarios. The volatile-to-refractory ratios, [O/Fe] = $1.25_{-0.74}^{+0.99} \mathrm{dex}$ and [C/Fe] = $0.60_{-0.74}^{+0.62} \mathrm{dex}$, fall within the solar to supersolar range. However, their large dispersions mean they can only provide tentative indications of volatile enrichment. Overall, the statistical significance of these constraints remains limited, making firm conclusions about the planet’s formation history premature. In the future, the combination of higher-quality high-resolution optical-to-infrared observations and \textit{JWST} data will enable more precise constraints on elemental abundances, providing more reliable insights into the formation and migration scenarios of UHJs. Finally, we advocate a retrieval-guided cross-correlation strategy to mitigate the risk of overlooking marginal species, exemplified by the tentative inference of CO in this study.
}


\maketitle

\section{Introduction}

Ultra-hot Jupiters (UHJs) are gas giant planets that orbit extremely close to their host stars, resulting in high dayside temperatures \citep[$T_\mathrm{day} \ge 2200\,\mathrm{K}$;][]{Parmentier+etal+2018}. Theoretical studies suggest that the atmospheres of these planets are dominated by atoms and ions rather than molecules due to extensive dissociation and ionization \citep{Lothringer+etal+2018,Parmentier+etal+2018,Arcangeli+etal+2018,Kitzmann+etal+2018}. The dayside atmospheres of UHJs consistently exhibit thermal inversions, driven by the presence of strong optical absorbers such as titanium oxide (TiO), vanadium oxide (VO), iron (Fe), and H$^-$ \citep{Hubeny+etal+2003,Fortney+etal+2008,Arcangeli+etal+2018,Lothringer+etal+2018,Parmentier+etal+2018,Gandhi+etal+2019}. 

In recent years, an increasing number of atmospheric chemical components in UHJs have been identified. These include neutral atoms with strong optical absorption features (such as \ion{H}{i}, \ion{Na}{i}, \ion{K}{i}, \ion{Ca}{i}, \ion{Fe}{i}, \ion{Si}{i}, \ion{Ti}{i}, and \ion{V}{i}), lower-order ions \citep[such as \ion{Ca}{ii}, \ion{Fe}{ii}, \ion{Mg}{ii}, and \ion{Ti}{ii}. e.g.][]{Yan+etal+2018,Hoeijmakers+etal+2019,D'Arpa+etal+2024,Borsato+etal+2023,Borsato+etal+2024,Casasayas-Barris+etal+2019,Nugroho+etal+2020b,Cont+etal+2022,Yang+etal+2024a,Kesseli+etal+2022,Pelletier+etal+2023,Prinoth+etal+2025,Prinoth+etal+2023}, and molecules with prominent infrared features, including CO, H$_2$O, OH, and TiO \citep[e.g.][]{Nugroho+etal+2021,Cont+etal+2021,Landman+etal+2021,Fu+etal+2022,Prinoth+etal+2022,vanSluijs+etal+2023}. The atmospheric chemical composition not only traces the dynamical processes within planetary atmospheres but also provides insights into their formation and migration history by constraining parameters such as metallicity and the [C/O] \citep[e.g.][]{Oberg+etal+2011,Madhusudhan+etal+2014,Madhusudhan+etal+2017,Mordasini+etal+2016}.

The ultra-high dayside temperatures of UHJs make them ideal targets for thermal emission observations. Dayside thermal emission observations measure the thermal radiation emitted from the daysides of UHJs. For instance, \ion{Fe}{i} emission lines have been reported on the daysides of several UHJs, including KELT-9b, WASP-189b, WASP-33b, and KELT-20b \citep{Pino+etal+2020,Kasper+etal+2021,Yan+etal+2020,Yan+etal+2022a,Nugroho+etal+2020a,Borsa+etal+2022,Johnson+etal+2023,Kasper+etal+2023}. CO emission lines have also been detected in many UHJs \citep[e.g. WASP-18b, WASP-76b, WASP-33b, and MASCARA-1b; see][]{Yan+etal+2023,vanSluijs+etal+2023,HolmbergandMadhusudhan2022}. Similarly, OH emission lines have been identified in three UHJs, namely WASP-33b, WASP-18b, and WASP-76b \citep[see][]{Nugroho+etal+2021,Brogi+etal+2023,Gandhi+etal+2024,WeinerMansfield+etal+2024}. \citet{Cont+etal+2021} further reported TiO emission lines on the dayside of WASP-33b. Furthermore, thermal emission spectroscopy is a powerful tool for identifying potential thermal inversions in UHJs, as it is highly sensitive to their thermal structures \citep[e.g.][]{Lothringer+etal+2018,Kitzmann+etal+2018,Helling+etal+2019,Fossati+etal+2021}.

In this paper, we present the detection of OH emission signals and the confirmation of Fe emission in the UHJ KELT-9b using the high-resolution ground-based near-infrared instrument SpectroPolarimètre InfraRouge \citep[SPIRou;][]{Donati+etal+2020}. KELT-9b \citep{Gaudi+etal+2017} is the hottest known UHJ, with an equilibrium temperature of approximately 4,000 K, orbiting an early A-type star with an orbital period of 1.48 days. The dayside atmosphere of KELT-9b has been extensively studied through thermal emission and transmission spectroscopy. \citet{Yan+etal+2018} first reported the detection of H$\alpha$ absorption in this planet's atmosphere using transmission spectroscopy. \citet{Mansfield+etal+2020} presented the Spitzer 4.5 $\mu$m phase curve of KELT-9b, finding evidence of H$_2$-H dissociation and recombination, which affect heat transport in the atmosphere \citep{BellandCowan2018,KomacekandTan2018,TanandKomacek2019}. \citet{Wyttenbach+etal+2020} reported the detection of the hydrogen Balmer series and constrained the planet's atmospheric mass loss rate. Over the past few years, numerous atoms and ions have been detected in the atmosphere of KELT-9b using thermal emission and transmission spectroscopy \citep{Hoeijmakers+etal+2018,Hoeijmakers+etal+2019,Yan+etal+2019,Borsa+etal+2021,Kasper+etal+2021,PaiAsnodkar+etal+2022,Sanchez-Lopez+etal+2022,Pino+etal+2022,Lowson+etal+2023,Ridden-Harper+etal+2023,Borsato+etal+2023,Borsato+etal+2024,D'Arpa+etal+2024}. However, no molecular components have been definitively identified in the atmosphere of KELT-9b so far, mainly due to the thermal dissociation of molecules under its extreme temperatures. Our detection of OH emission signals, therefore, not only confirms the presence of a thermal inversion in the dayside atmosphere of KELT-9b but also provides valuable insights into the chemical composition and physical processes governing the atmosphere of this extreme UHJ.

This work is organized as follows. Sects.~\ref{OBS} and~\ref{DR} describe the observations and data reduction processes. In Sect.~\ref{CCF}, we present the detection of atmospheric components using cross-correlation techniques applied to SPIRou data, followed by a discussion of the results. Sect.~\ref{RETRIEVAL} details the methodology and outcomes of the atmospheric retrievals. Sect.\ref{RG} introduces a retrieval-guided cross-correlation strategy that leverages atmospheric retrieval results to inform cross-correlation analyses and improve species detectability. Finally, the conclusions are summarized in Sect.~\ref{CONCL}.

\section{Observations}\label{OBS} 

We obtained two observations of the dayside of KELT-9b using SPIRou \citep{Donati+etal+2020}. SPIRou is a high-resolution ($R\sim70,000$) cryogenic fiber-fed echelle spectro-polarimeter with broad wavelength coverage from 0.95 to 2.5 $\mu$m across 49 spectral orders. It is mounted on the 3.6 m Canada-France-Hawaii Telescope (CFHT). The incoming light from the target is split into two orthogonal polarization states by an achromatic polarimeter and then separately fed into two science fibers (fiber A and B). A third fiber (fiber C) is used to inject the signal from a calibration source. SPIRou's simultaneous wavelength coverage and high spectral resolving power in the near-infrared make it an ideal instrument for studying molecules such as H$_2$O, OH, and CO.

\begin{figure}
\centering
\includegraphics[trim=0cm 0.0cm 0.cm 0.cm, clip=true,width=\hsize]{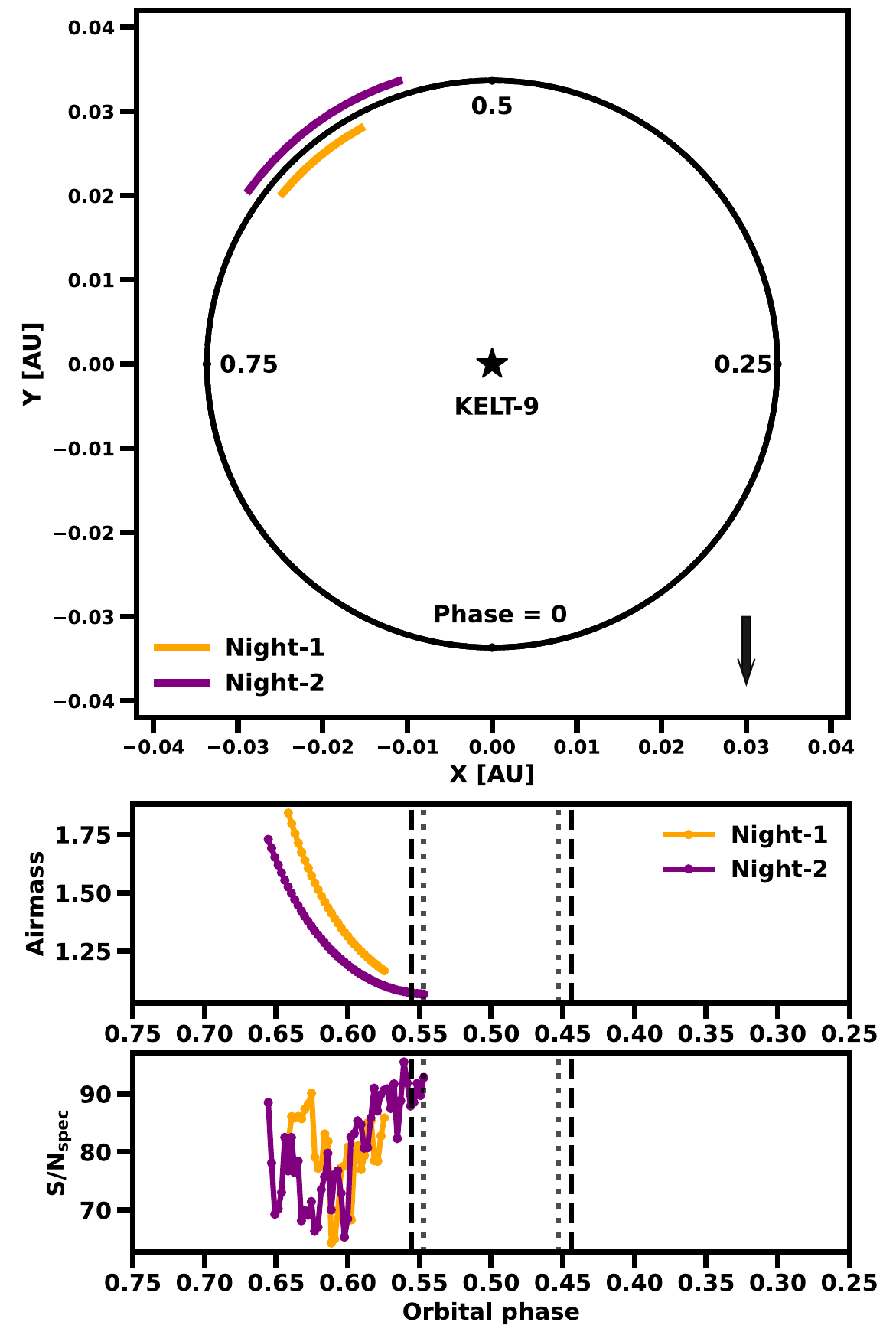}
\caption{\emph{Upper panel}: Phase coverage of the thermal emission observations. \emph{Middle panel}: Airmass as a function of orbital phase for each observation. \emph{Lower panel}: S/N$_\mathrm{spec}$ in the spectral order of the He infrared triplet (10830 \AA) as a function of orbital phase for each observation. The colours represent different observation dates. The dashed and dotted lines indicate the eclipse duration ($T_\mathrm{14}$) and the fully in-eclipse duration ($T_\mathrm{23}$), respectively.}
\label{ObsLog}
\end{figure}

Two observations were conducted at orbital phases near the secondary eclipse to capture the thermal emission signal from the planet. A summary of the observation log is provided in Table~\ref{ObservationInfo}, while Fig.~\ref{ObsLog} illustrates the orbital phases of the observations along with the airmass and signal-to-noise ratio of the observed stellar spectra (S/N$_\mathrm{spec}$). The first four exposures from Night-2, which occurred during the secondary eclipse phases, were discarded.

\begin{table*}[htb!]
\caption{Summary of the observation logs.}             
\label{ObservationInfo}      
\centering          
\begin{tabular}{l c c c c c c c c}     
\hline\hline 
\noalign{\smallskip}      
Instrument & Night & UT Date & UT Time & Exposure Time [s] & $N_\text{obs}$ & Program ID & PI\\
\noalign{\smallskip}
\hline\noalign{\smallskip}                  
    SPIRou & Night-1 & 2020-07-28 & 12:22$-$14:45 & 273 & 30 & 20AS17 & Hui Zhang\\  
    SPIRou & Night-2 & 2020-07-31 & 10:29$-$14:20 & 273 & 48 & 20AS17 & Hui Zhang\\
\noalign{\smallskip}
\hline                 
\end{tabular}
\end{table*}

\section{Data reduction}\label{DR}

The raw data were reduced using the SPIRou data reduction pipeline \texttt{APERO} \citep[A PipelinE to Reduce Observations;][]{Cook+etal+2022}. The \texttt{APERO} pipeline produces order-by-order science spectra, which include the flux from each of the two science fibers individually (labelled as \texttt{FluxA} and \texttt{FluxB}) and their combined flux (labelled as \texttt{FluxAB}). For our analysis, we used only the \texttt{FluxAB} format, as polarization information is not required for detecting molecules in exoplanet atmospheres. Additionally, the pipeline provides telluric-corrected spectra, derived from the Earth's transmittance reconstructed using a library of SPIRou telluric observations. However, we used the spectra without telluric correction in our analysis, as the empirical telluric correction may introduce additional noise.

We first normalized the spectrum for each order of each reduced spectrum using the blaze function provided by the SPIRou pipeline. To identify outliers, we flagged data points that deviated by more than five standard deviations from the smoothed spectrum obtained by applying a median filter with a window size of nine pixels. These flagged points were replaced with the median of the surrounding thirty data points. Furthermore, we remove continuum variations following the method of \citet{Chen+etal+2020}, using a fourth-order polynomial to fit the residuals between each individual spectrum and the reference spectrum. We empirically derived noise from the pipeline's raw flux outputs and accounted for them in all subsequent analyses.

\begin{figure}
\centering
\includegraphics[trim=0.cm 0cm 0cm 0.cm, clip=true,width=\hsize]{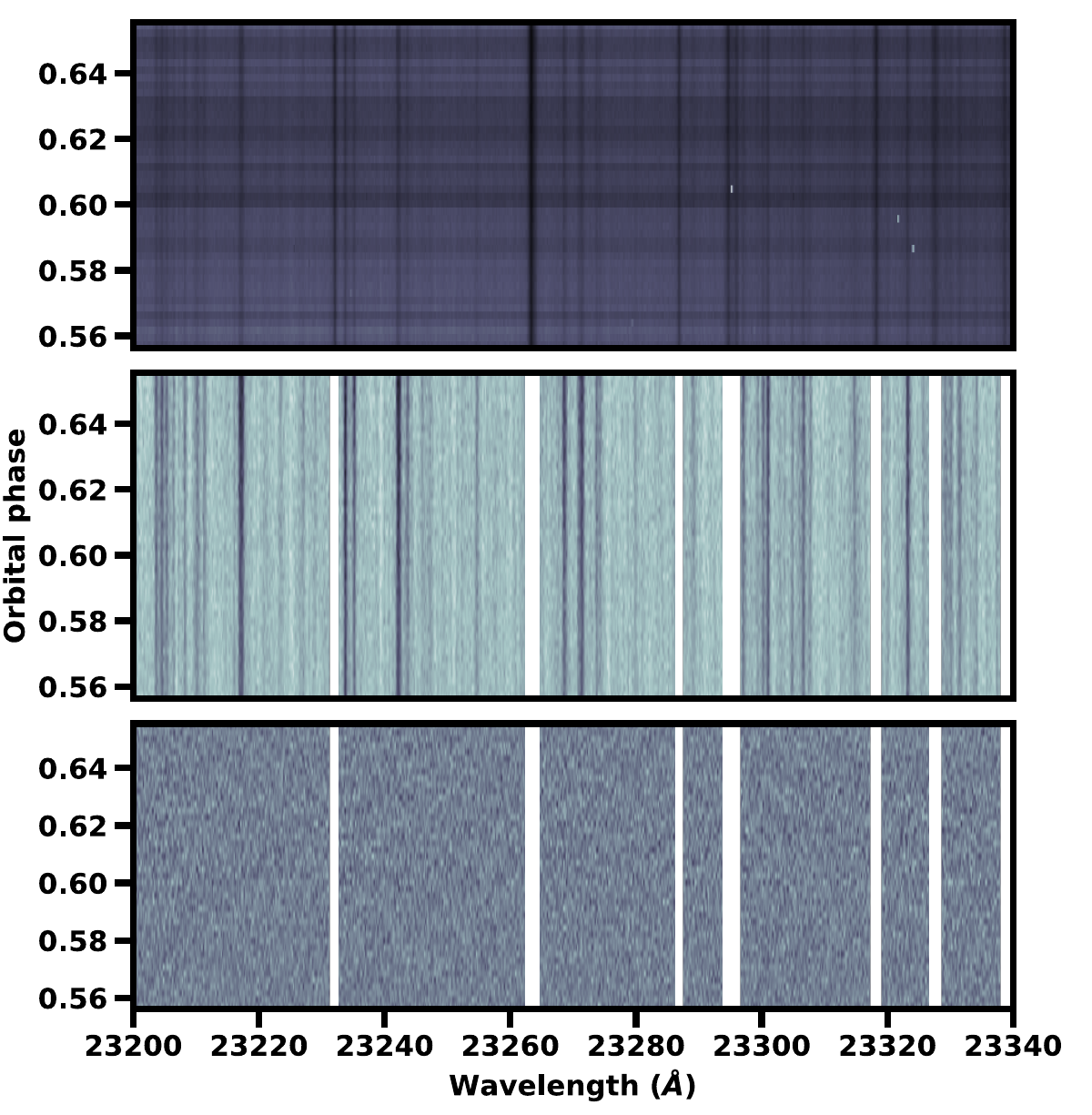}
\caption{Example of data reduction steps. \emph{Top panel}: Raw spectral matrix. \emph{Middle panel}: Spectral matrix after normalization, outlier correction, and masking of the strongest telluric features and lowest S/N$_\mathrm{spec}$ regions. \emph{Bottom panel}: Spectral matrix after SYSREM treatment. Telluric and stellar spectral lines have been removed by the SYSREM algorithm, leaving planetary signals buried in this residual spectral matrix.}
\label{DRS}
\end{figure}

We adopted the method detailed by \citet{Yang+etal+2024a} and \citet{Yang+etal+2024b} for data reduction. The spectral matrix was constructed by collecting the cleaned, normalized spectra from each observation, sorted by time. Due to saturation absorption in the Earth's atmosphere at certain near-infrared wavelengths, the spectral orders within these wavelength ranges were discarded. The wavelength bins with the lowest 5\% S/N$_\mathrm{spec}$ in each order were also masked. To minimize contributions from telluric and stellar lines in each spectral order, we applied the SYSREM algorithm \citep{Tamuz+etal+2005}. This algorithm fits each wavelength bin in the spectral matrix to capture systematics caused by telluric contamination, stellar lines, and instrumental effects, and then removes these systematics from the spectral matrix. The empirically derived noise is estimated from the SYSREM-cleaned residuals by calculating the standard deviations along both the wavelength and time axes and combining them in quadrature. After several iterations, SYSREM produces a residual spectral matrix that retains only planetary features. We followed the approach proposed by \citet{Gibson+etal+2020}, dividing the spectral matrix by the systematics instead of subtracting them. Figure~\ref{DRS} illustrates an example of the data reduction steps for a single night.

\section{Detection of OH and Fe emission lines}\label{CCF}

\subsection{Planetary atmosphere model spectra}\label{MS}

To search for atmospheric species using the cross-correlation technique, we first construct model spectra for the planetary atmosphere and then cross-correlate these model spectra with the residual spectra obtained from the SYSREM reduction. The \texttt{petitRADTRANS} code \citep[v2.7.7;][]{Molliere+etal+2019} is used to calculate the thermal emission model spectra for each species.

To construct the model, we assumed different constant abundance grids for various species instead of calculating the mean molecular weight and volume mixing ratios (VMRs) from equilibrium chemistry and solar elemental abundances. The abundance grids were not based on any specific chemical scenario, but were chosen to maximize the significance of the detection without imposing assumptions about atmospheric composition. Following \citet{Brogi+etal+2014} and \citet{Yan+etal+2020}, we adopted a two-point parameterized temperature-pressure (T-P) profile, adapted from the retrieval results of \citet{Kasper+etal+2021}. Table~\ref{WASP76INFO} lists the planetary parameters we used, such as the radius and surface gravity. The opacities of the chemical species used to calculate the model spectra were provided by \texttt{petitRADTRANS} with its precalculated line lists\footnote{\url{https://petitradtrans.readthedocs.io/en/2.7.7/content/available_opacities.html}}, except for CO. For CO, we used the line list from \citet{Li+etal+2015}, stored in the ExoMol database, as it covers a wider temperature range, making it more suitable for high-temperature atmospheres.

The model spectra were convolved with the instrument profile to match the resolution of the instrument before being used in the cross-correlation analysis. The model spectra were first normalized by dividing by the blackbody spectrum corresponding to the bottom atmospheric temperature and converted into the planet–to–star flux ratio. After these steps, the model spectra were normalized to have values close to one, making them suitable for the cross-correlation analysis.

\subsection{Cross-correlation}

We independently performed cross-correlation \citep{Snellen+etal+2010} for each molecular species to search for evidence of molecules in the planetary atmosphere. The model spectra were shifted from $-$200 to +200 km s$^{-1}$ in steps of 2 km s$^{-1}$, approximately matching the sampling step per pixel of SPIRou. At each shift, the weighted residual spectrum was multiplied by the shifted model spectrum to compute a weighted cross-correlation function (CCF). The CCF is defined as
\begin{equation}
\mathrm{CCF}=\sum \frac{r_{i} m_{i}(\varv)}{\sigma_{i}^{2}},
\end{equation}
where $r_{i}$ is the residual spectrum, $m_{i}$ is the model spectrum shifted by the velocity $\varv$, and $\sigma_{i}$ is the error at the wavelength point $i$. After calculating the CCFs for each residual spectrum, we stacked all the CCFs to generate a CCF map for each observation night and model spectrum.

The CCF map was converted from the Earth rest frame to the planet rest frame using a grid of assumed planetary semi-amplitudes ($K_\mathrm{p}$). Assuming a circular orbit for the planet, the radial velocity (RV) of the planet $\varv_\mathrm{p}$ is defined as
\begin{equation}\label{FRV}
\varv_{\mathrm{p}}=\varv_\mathrm{sys}-\varv_\mathrm{bary}+K_\mathrm{p} \sin (2 \pi \phi)+\Delta \varv,
\end{equation}
where $\varv_\mathrm{sys}$ is the systemic velocity, $\varv_\mathrm{bary}$ is the barycentric Earth radial velocity (i.e. BERV), $\Delta \varv$ is the radial velocity deviation from zero, and $\phi$ is the orbital phase. In general, $\Delta \varv$ is closely related to the dynamical features of the planetary atmosphere (e.g. day-to-night winds). A phase-folded one-dimensional CCF was generated for each assumed value of $K_\mathrm{p}$ by averaging the CCF map, which was shifted into the planetary rest frame based on the assumed $K_\mathrm{p}$. By stacking the one-dimensional CCFs for different $K_\text{p}$ values ranging from 0 to 300 km s$^{-1}$ with a step of 1 km s$^{-1}$, we created a $K_\mathrm{p}$ map. 
The $K_\mathrm{p}$ map was normalized by dividing it by the standard deviation ($\sigma$) obtained from a Gaussian fit to the distribution of all $K_\mathrm{p}$ values. As a result, the final $K_\mathrm{p}$ map was expressed in terms of significance value (S/N). Uncertainties on $K_\mathrm{p}$ and $\Delta \varv$ were estimated as the parameter offsets corresponding to a decrease of one in S/N from its peak value on the $K_\mathrm{p}$ map.

\subsection{Cross-correlation results}

\begin{figure*}
\centering
\includegraphics[width=\hsize]{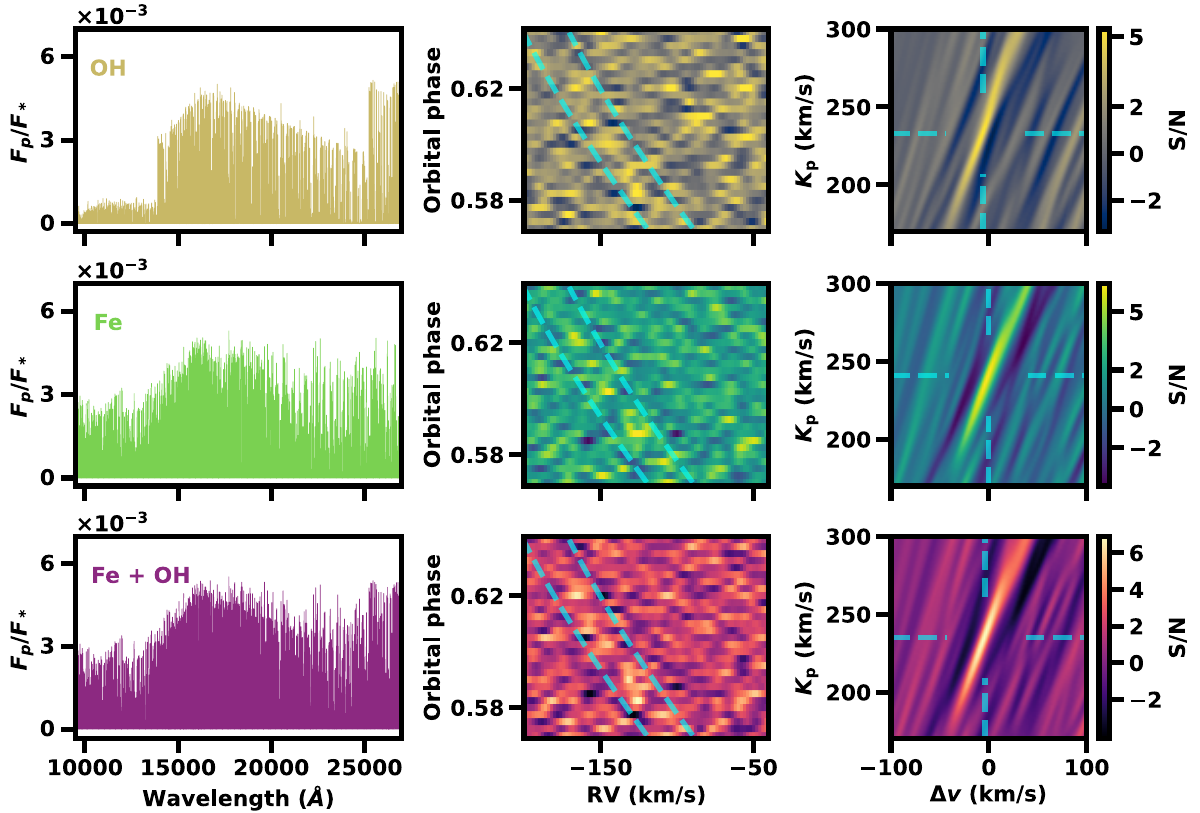}
\caption{Model spectra, CCF maps, and $K_\mathrm{p}$ maps for OH, Fe, and both species combined. \emph{Left panel} shows the model spectra after continuum removal. \emph{Middle panel} displays the CCF map in the stellar rest frame (i.e. corrected for BERV and systemic velocity). The dashed cyan line represents the planetary RV derived from the expected value of $K_\mathrm{p}$. \emph{Right panel} presents the $K_\mathrm{p}$ map for each molecule. The dashed lines intersecting at the cross mark indicate the position of the S/N peak. The colour bar represents the S/N values.}
\label{CCF_FeOH}
\end{figure*}

\begin{figure}
\centering
\includegraphics[trim=0.7cm 0.4cm 0.7cm 0.45cm, clip=true, width=\hsize]{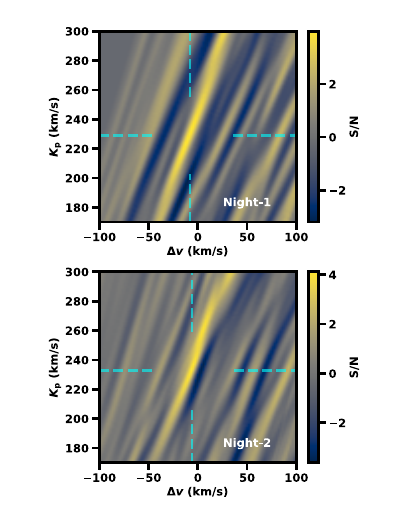}
\caption{$K_\mathrm{p}$ map of OH for each individual night. The format is the same as the right panel of Fig.~\ref{CCF_FeOH}.}
\label{CCF_OH_ind}
\end{figure}

We applied the SYSREM algorithm over ten iterations and adopted the residual spectral matrix from the iteration that yielded the highest S/N in the $K_\mathrm{p}$ map as our final result. Using the cross-correlation method, we detected OH and Fe signals in the dayside atmosphere of KELT-9b. This marks the first detection of molecular species in KELT-9b. The combined CCF map and $K_\mathrm{p}$ map for these species over two nights are shown in Fig.~\ref{CCF_FeOH}. The model spectra shown in Fig.~\ref{CCF_FeOH} assume abundances of $10^{-3}$ for OH and $10^{-5}$ for Fe, respectively. The OH emission signal was detected with an S/N of 5.3 at $K_\mathrm{p} = 233_{-4}^{+5}$ km s$^{-1}$ and $\Delta \varv = -6_{-4}^{+4}$ km s$^{-1}$ by combining the two observations. The $K_\mathrm{p}$ map of the OH signal for each night is shown in Fig.~\ref{CCF_OH_ind}. For Night-1, the OH signal has an S/N of 4.0 at $K_\mathrm{p} = 229_{-6}^{+6}$ km s$^{-1}$ and $\Delta \varv = -8_{-4}^{+4}$ km s$^{-1}$. For Night-2, the S/N is 4.1 at $K_\mathrm{p} = 233_{-4}^{+7}$ km s$^{-1}$ and $\Delta \varv = -6_{-4}^{+4}$ km s$^{-1}$. The location of maximum S/N aligns well with the expected radial velocity (RV) of the planet's orbital motion, inferred from its orbital parameters. Additionally, we confirmed the presence of Fe in the dayside atmosphere of KELT-9b, which had been previously detected using optical instruments \citep[e.g.][]{Pino+etal+2020,Kasper+etal+2021}. The detections of OH and Fe emission with SPIRou's near-infrared data confirm the existence of thermal inversion layers in KELT-9b's upper atmosphere, as reported in multiple previous optical studies.

Furthermore, we searched for other species (e.g. H$_2$O, CO, TiO, VO, FeH, \ion{Fe}{ii}, Ti, and V) but did not detect any significant signals in the SPIRou data by CCF method. The detection of OH, together with no clear detection of H$_2$O, suggests that H$_2$O undergoes thermal dissociation due to the high temperature on the dayside \citep{Parmentier+etal+2018,Arcangeli+etal+2018,Lothringer+etal+2018}. The non-detection of other molecules and metal atoms and ions could be attributed to SPIRou's wavelength coverage (0.95–2.5 $\mu$m), which does not encompass a significant number of spectral lines for these species, as they are primarily abundant in optical bands.

\section{Retrieval of the planetary atmospheric properties}\label{RETRIEVAL}

\subsection{Retrieval method}

Retrieval methods are used to derive planetary atmospheric parameters by comparing parameterized planetary models with observational data through Bayesian inference. In recent years, planetary atmospheric retrieval techniques for high-resolution spectroscopy have advanced rapidly \citep[e.g.][]{Brogi+etal+2019,Gibson+etal+2020,Yan+etal+2020}. In this work, we adopt the retrieval framework developed by \citet{Yan+etal+2022a,Yan+etal+2023} and \citet{Lesjak+etal+2023} to characterize the atmospheric properties of the dayside of KELT-9b.

We used the one-dimensional radiative transfer code \texttt{petitRADTRANS} \citep{Molliere+etal+2019} to forward model the thermal emission spectrum of KELT-9b. Our calculations included the two CCF‐detected species (OH and Fe), as well as other high-abundance species (VMR $>10^{-7}$ at pressures less than $1\,\mathrm{bar}$) predicted by chemical equilibrium and exhibiting strong cross sections in the instrument's wavelength range (CO and H$_2$O). This selection provides a relatively complete opacity set, and potential missing contributors are less concerning in this case. In particular, TiO is unlikely to be present at the high temperature of KELT-9b, and the relatively red wavelength coverage of the spectrograph avoids severe crowding from metal lines. We then described the atmosphere's thermal structure using a two-point $T$-$P$ parameterized profile. The model also incorporated the effect of H$^{-}$, an important continuum opacity that mutes the thermal emission spectral features of UHJs \citep{Arcangeli+etal+2018,Lothringer+etal+2018,Parmentier+etal+2018}. The atmosphere was divided into 25 layers, uniformly distributed on a logarithmic scale between 10$^{-6}$ bar and 100 bar, representing a trade-off between computational efficiency and the accuracy of the calculated spectra. We set the low-altitude temperature $T_1$, low-altitude pressure $P_1$, high-altitude temperature $T_2$, and high-altitude pressure $P_2$ as free parameters in the $T$-$P$ profile. The mixing ratios of all species were computed on-the-fly by \texttt{FastChem} \citep{Stock+etal+2018} under chemical equilibrium. Given each trial two-point $T$–$P$ profile and the specified [M/H] and elemental abundance ([Fe/H], [O/H], and [C/H]; that is, the logarithm of the ratio between the certain element-to-hydrogen ratio of the planet and the certain element-to-hydrogen ratio of the Sun), \texttt{FastChem} calculated the equilibrium abundances of the species in real time. The model spectra were expressed as 1+$F_\mathrm{p}$/$F_\mathrm{s}$, where $F_\mathrm{p}$ was computed using \texttt{petitRADTRANS} and $F_\mathrm{s}$ was assumed to be a blackbody spectrum at the stellar effective temperature (9329 K). To account for line profile broadening, including instrumental effects and planetary rotation, the model spectra were convolved with the instrument profile (assumed to be a Gaussian corresponding to SPIRou's resolution) and the rotational broadening kernel, following the method in \citet{Carvalho+etal+2023}, originally from \citet{Gray+etal+1992}, which has been widely used for studying stellar rotation. We assumed a linear limb-darkening law and fixed the limb-darkening coefficient $\epsilon = 1$, meaning that radiation from the limb does not contribute to the total flux. Additionally, we assumed that the inclination of the planetary equator matches the orbital inclination, consistent with the hypothesis of likely tidally locked rotation.

We then Doppler-shifted the model spectra to generate a model spectral matrix with the same dimensions and size as the residual spectral matrix. Since the residual spectral matrix is in the observer's rest frame, each model spectrum was shifted according to the radial velocity calculated using Eq.~\ref{FRV}, depending on the values of $K_\mathrm{p}$ and $\Delta \varv$. Next, each model spectrum was interpolated onto the wavelength grid of the observed residual spectrum, and the resulting one-dimensional model spectra were arranged into a two-dimensional matrix with the same shape as the residual spectral matrix.

Before fitting the model spectral matrix to the residual spectral matrix, it is essential to ensure that the model and observational data are processed in the same way. Since the observational data were reduced using SYSREM, which alters the strength and profile of spectral lines, the model spectrum must also be processed through SYSREM. To improve computational efficiency, we adopted a fast SYSREM filtering method following \citet{Yan+etal+2023} and developed by \citet{Gibson+etal+2022}. The SYSREM filter matrix was generated during the application of SYSREM to the observational spectral matrix. This filter matrix was then applied to the model spectral matrix, resulting in the final filtered model matrix for retrieval. Finally, a Gaussian high-pass filter with a Gaussian $\sigma$ of 21 points was applied to both the final model matrix and the residual matrix to remove any remaining broadband features.

We employed a standard Gaussian log-likelihood function
\begin{equation}\label{Likefun}
\ln L=-\frac{1}{2} \sum_{i, j}\left[\frac{\left(R_{i j}-M_{i j}\right)^{2}}{\left(\beta \sigma_{i j}\right)^{2}}+\ln 2 \pi\left(\beta \sigma_{i j}\right)^{2}\right]
\end{equation}
to evaluate the likelihood of the model spectral matrix relative to the residual spectral matrix. In this expression, $R_{i j}$ represents the residual spectral matrix at the wavelength point $i$ and the time (phase) $j$, $M_{i j}$ represents the model spectral matrix, $\sigma_{i j}$ denotes the uncertainties of the residual spectra, and $\beta$ is the scale factor of these uncertainties. We used the residual spectral matrix corresponding to the SYSREM iteration that yielded the maximum S/N for the detection of the OH and Fe lines. The log-likelihood function was evaluated using Markov Chain Monte Carlo (MCMC) sampling implemented with the \texttt{emcee} package \citep{Foreman-Mackey+etal+2013}. The free parameters and their boundaries are listed in Table~\ref{FreeParam}. While $\beta$ was assumed to follow a normal distribution, all other parameters were assumed to follow uniform distributions. The MCMC simulation was run with 20,000 steps and 32 walkers for each free parameter.

\begin{table}
\caption{Best-fit parameters from the atmospheric retrieval on KELT-9b.}
\label{FreeParam}
\centering
\renewcommand{\arraystretch}{1.4}
\begin{threeparttable}
    \begin{tabular}{l c c c}
        \hline\hline
        \multirow{2}*{Parameter} & Chemical equilibrium  & \multirow{2}*{Prior} & \multirow{2}*{Unit}  \\     
                                 & retrieval             &                      &                       \\    
        \hline
        \noalign{\smallskip}
        $T_2$                           & $5608_{-602}^{+1119}$          & $\mathcal{U}(1000,8100)$    & K              \\
        $T_1$                           & $1916_{-663}^{+1224}$           & $\mathcal{U}(1000,8100)$    & K              \\
        $\log{p_2}$                     & $-1.8_{-0.9}^{+0.6}$           & $\mathcal{U}(-4,-1)$        & log\,bar       \\
        $\log{p_1}$                     & $-0.2_{-0.6}^{+1.2}$           & $\mathcal{U}(-1,2)$         & log\,bar       \\
        $\Delta \varv$                  & $-2.9_{-2.9}^{+2.8}$           & $\mathcal{U}(-15,15)$       & km\,s$^{-1}$   \\
        $K_\mathrm{p}$                  & $236.0_{-4.3}^{+4.7}$          & $\mathcal{U}(204,264)$      & km\,s$^{-1}$   \\
        $\varv_\mathrm{eq}$             & $7.9_{-2.0}^{+2.2}$            & $\mathcal{U}(0,20)$         & km\,s$^{-1}$   \\
        $\beta$                         & $0.433 \pm 0.0001$             & $\mathcal{N}(1,0.5)$        & \ldots         \\
        $\mathrm{[Fe/H]}$               & $-0.58_{-0.22}^{+0.26}$         & $\mathcal{U}(-3,3)$        & dex            \\
        $\mathrm{[O/H]}$                & $0.61_{-0.58}^{+1.19}$         & $\mathcal{U}(-3,3)$         & dex            \\
        $\mathrm{[C/H]}$                & $0.07_{-0.67}^{+0.72}$         & $\mathcal{U}(-3,3)$         & dex            \\
        $\mathrm{[M/H]}$                & $-1.17_{-1.31}^{+1.64}$        & $\mathcal{U}(-3,3)$         & dex            \\
        \noalign{\smallskip}
        \hline
    \end{tabular}
\end{threeparttable}
\end{table}

\subsection{Retrieval results and discussion}\label{Retrieval}

The retrieval was first performed using the combined observational data. The retrieved parameter values are listed in Table~\ref{FreeParam}, and their posterior distributions are shown in Fig.~\ref{Retri_OtoH_corner}. Additionally, we carried out independent chemical equilibrium retrievals for each observation separately. To further assess the robustness of our results, we also tested retrievals with different SYSREM iterations (see Fig.~\ref{Diff_iteration_marginals_compare} and Fig.~\ref{Diff_iteration_TP_compare}). The results indicate that the choice of iteration has limited effect on this dataset and does not significantly impact the overall conclusions.

\begin{figure}
\centering
\includegraphics[width=\hsize]{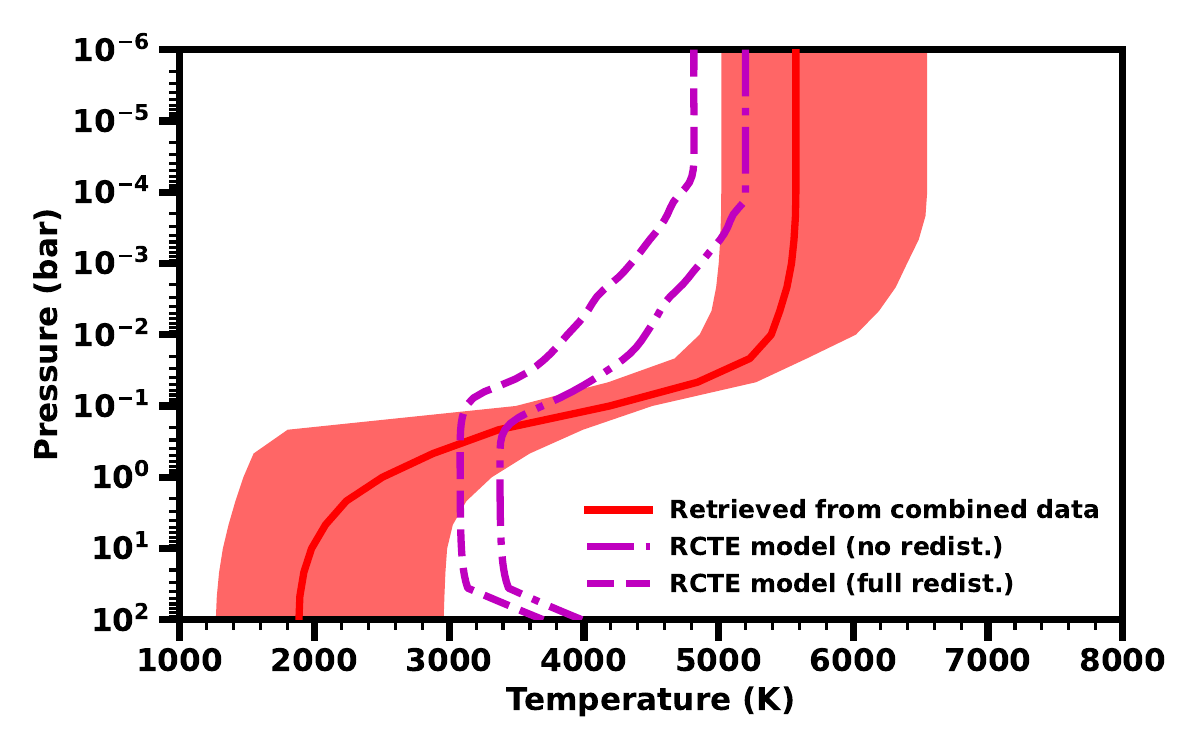}
\caption{Retrieved $T$-$P$ profiles from chemical equilibrium retrieval of combined two-night observations, compared with RCTE model profiles. The solid line represents the median, while the shaded region indicates the $1\sigma$ range of the sampled $T$-$P$ profiles. The RCTE models were calculated assuming solar abundances, with one model representing no heat redistribution from the dayside to the nightside and the other assuming full heat redistribution.}
\label{Retri_OtoH_TP}
\end{figure}

\begin{figure}
\centering
\includegraphics[width=\hsize]{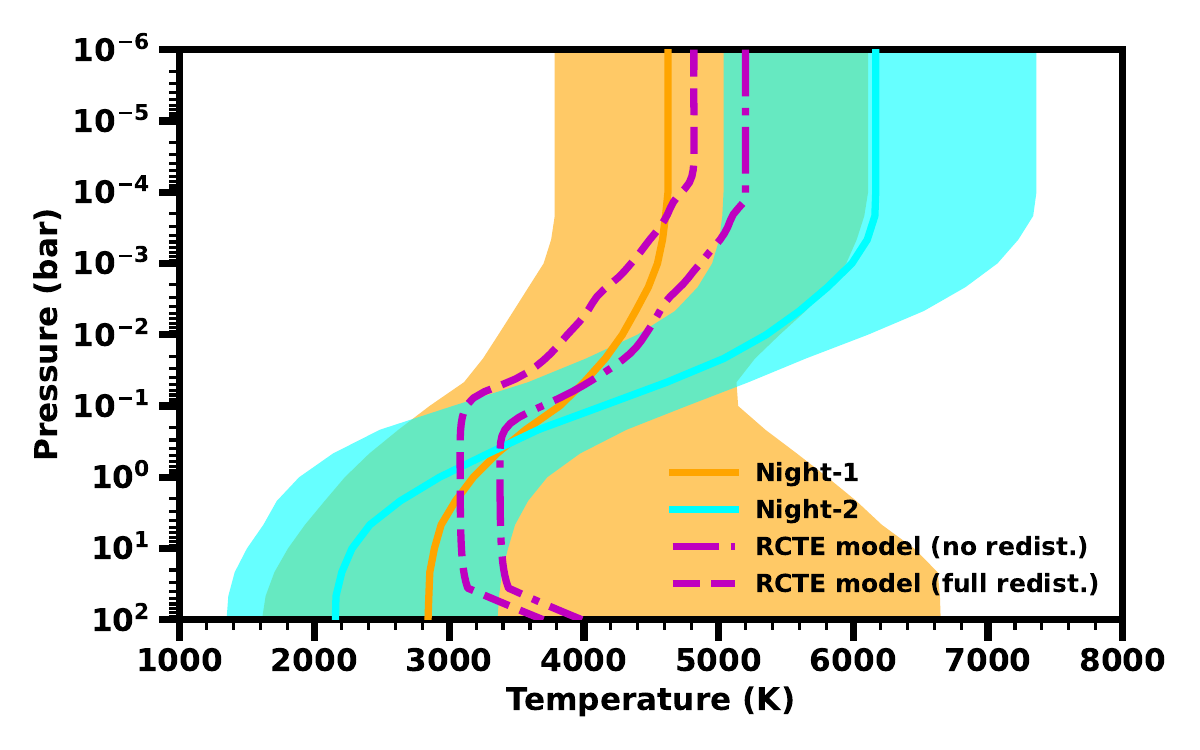}
\caption{Retrieved $T$-$P$ profiles for individual observational nights. Same format as Fig.~\ref{Retri_OtoH_TP}.}
\label{Retri_OtoH_TP_ind}
\end{figure}

The retrieved $T$-$P$ profile obtained from the retrieval using the combined observations is presented in Fig.~\ref{Retri_OtoH_TP}. This retrieval confirms the presence of a thermal inversion layer on the dayside of the atmosphere. The thermal inversion layer spans a pressure range from approximately $10^{-2}$ bar to 1 bar, with temperatures of $T_{2} = 5608_{-602}^{+1119}$ K in the upper atmosphere and $T_{1} = 1916_{-663}^{+1224}$ K in the lower atmosphere. The retrieved inversion layer is slightly deeper than previous results reported using optical observations \citep[e.g.][]{Kasper+etal+2021}, as infrared wavelengths probe deeper atmospheric layers compared to optical wavelengths. In previous studies, degeneracy between retrieved metallicity and pressure often arose because only metal atoms and ions were detected. Higher metallicity implies a more opaque upper atmosphere, shifting the optically thick layer to lower pressures and moving the thermal  inversion layer upwards. Conversely, lower metallicity shifts the inversion layer deeper into the atmosphere. In this work, the detection of OH in the near-infrared helps break this degeneracy, allowing for a more precise determination of the thermal inversion layer's location. Additionally, we performed a radiative-convective thermochemical equilibrium (RCTE) calculation for KELT-9b using PICASO \citep{Batalha+etal+2019,Mukherjee+etal+2023}. It should be noted that this model does not account for non-local thermodynamic equilibrium (NLTE) effects and excludes most ion opacities, which are known to significantly influence the thermal structure of the upper atmosphere of UHJs \citep{Fossati+etal+2020,Fossati+etal+2021}. Despite these limitations, the RCTE model remains consistent with the retrieved $T$-$P$ profile (see Fig.~\ref{Retri_OtoH_TP}). We infer that NLTE effects and the opacities of metal ions primarily govern heating and cooling in the upper atmosphere but have limited impact on the middle atmosphere within the pressure ranges probed by infrared bands. In the middle atmosphere, heating is dominated by metal atoms (e.g. Fe), TiO, and VO rather than metal ions. As shown in Fig.~\ref{Retri_OtoH_TP_ind}, the retrieved $T$-$P$ profiles for different observation nights are broadly consistent, likely due to the similar orbital phase coverage across the two nights. The retrieved bottom atmospheric temperatures for Night-1 are poorly constrained, which may be related to its lower data quality.

As followed by \citet{Yan+etal+2023}, we constrained the planet's equatorial rotation speed, $v_\mathrm{eq}$, to be $7.90_{-1.96}^{+2.18}$ km s$^{-1}$. These values are consistent with the speed expected for tidally locked rotation, which is $6.80$ km s$^{-1}$. Additionally, atmospheric dynamics, such as equatorial super-rotation jets, could induce a broadening of the line profile. However, our retrieval results show no significant additional line broadening beyond that caused by planetary rotation. We therefore conclude that this planet lacks significant equatorial super-rotation jets, which are predicted by general circulation models (GCMs) to be the dominant atmospheric dynamical structures in hot Jupiters.

The retrieval results indicate that $K_\mathrm{p}$ is approximately $236_{-4}^{+5}$ km s$^{-1}$, consistent with the expected value of $234$ km s$^{-1}$. The retrieved $\Delta \varv$ value is approximately $-2.9_{-2.9}^{+2.8}$ km s$^{-1}$, suggesting no significant blue- or red-shifted signals. Here, we do not aim to draw definitive conclusions about atmospheric dynamical signatures. First, there is substantial uncertainty in the systemic velocity of an early-type star like KELT-9b, with different values reported in the literature, such as $-17.74 \pm 0.11$ km s$^{-1}$ by \citet{Hoeijmakers+etal+2019}, $-19.819 \pm 0.024$ km s$^{-1}$ by \citet{Borsa+etal+2019}, and $-21.61 \pm 0.77$ km s$^{-1}$ by \citet{Stangret+etal+2024}. Second, unlike transmission spectroscopy, which has a fixed phase coverage, dayside thermal emission observations must account for the coupled effects of rotational and atmospheric dynamics at different phases. This requires considering visible hemispheric variations, which can lead to changes in the observed net Doppler shift. For instance, \citet{Stangret+etal+2024} and \citet{D'Arpa+etal+2024} reported evidence of day-to-night winds on KELT-9b based on transmission spectra. However, both our results and those of \citet{Kasper+etal+2021} found no significant net Doppler shifts in KELT-9b from dayside thermal emission spectra. To better interpret atmospheric dynamics signatures from net Doppler shifts in thermal emission spectroscopy, more detailed modelling studies and phase-resolved observations are needed.

\begin{figure}
\centering
\includegraphics[width=\hsize]{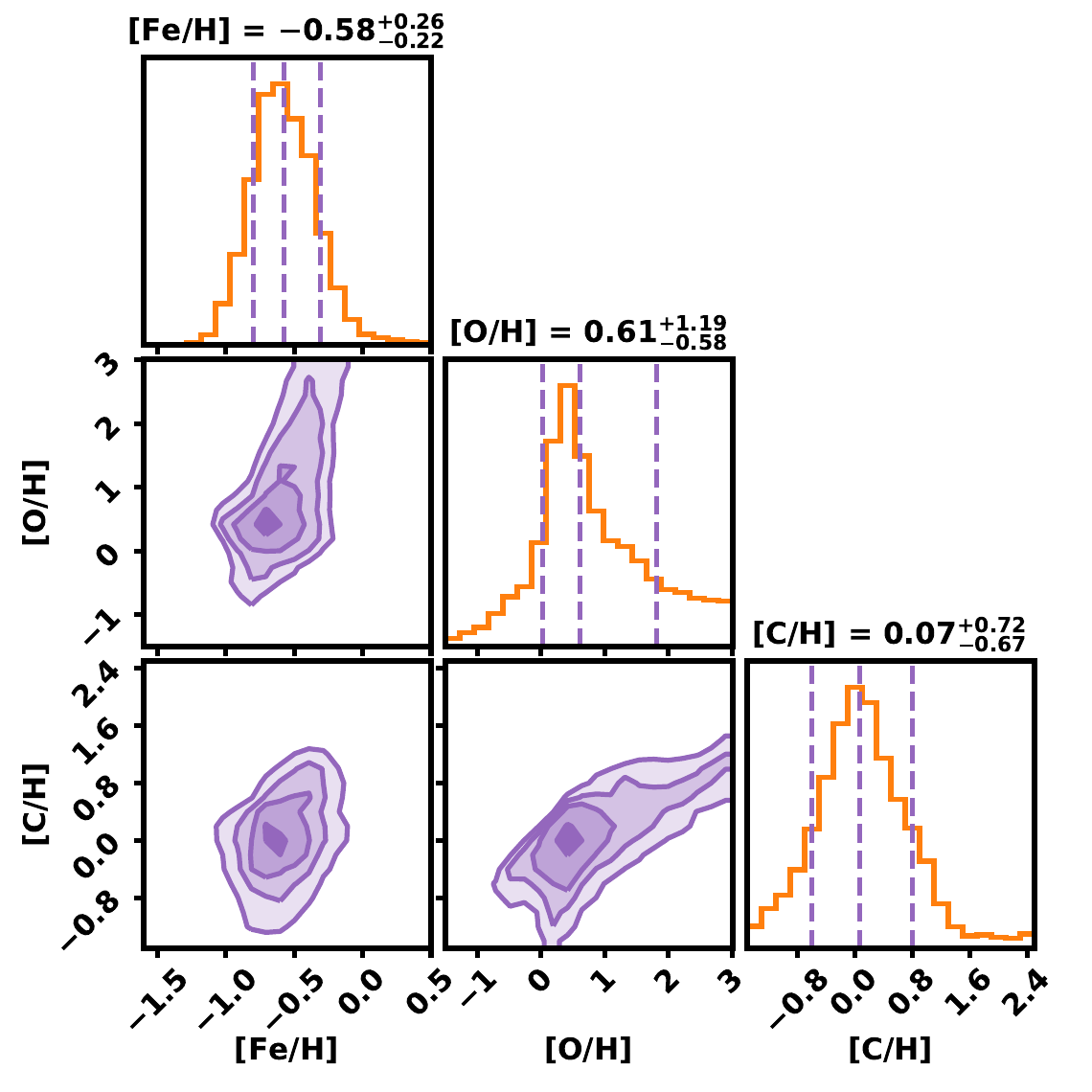}
\caption{
Retrieved logarithmic elemental abundances in KELT-9b's dayside atmosphere under chemical equilibrium. The corner plots show the posterior distributions and correlations between the atmospheric parameters. The dashed vertical lines in the posterior distributions denote the median and $\pm 1 \sigma$ credible intervals. All abundances are given relative to Sun.
}
\label{Retri_element_XH}
\end{figure}

\begin{figure}
\centering
\includegraphics[width=\hsize]{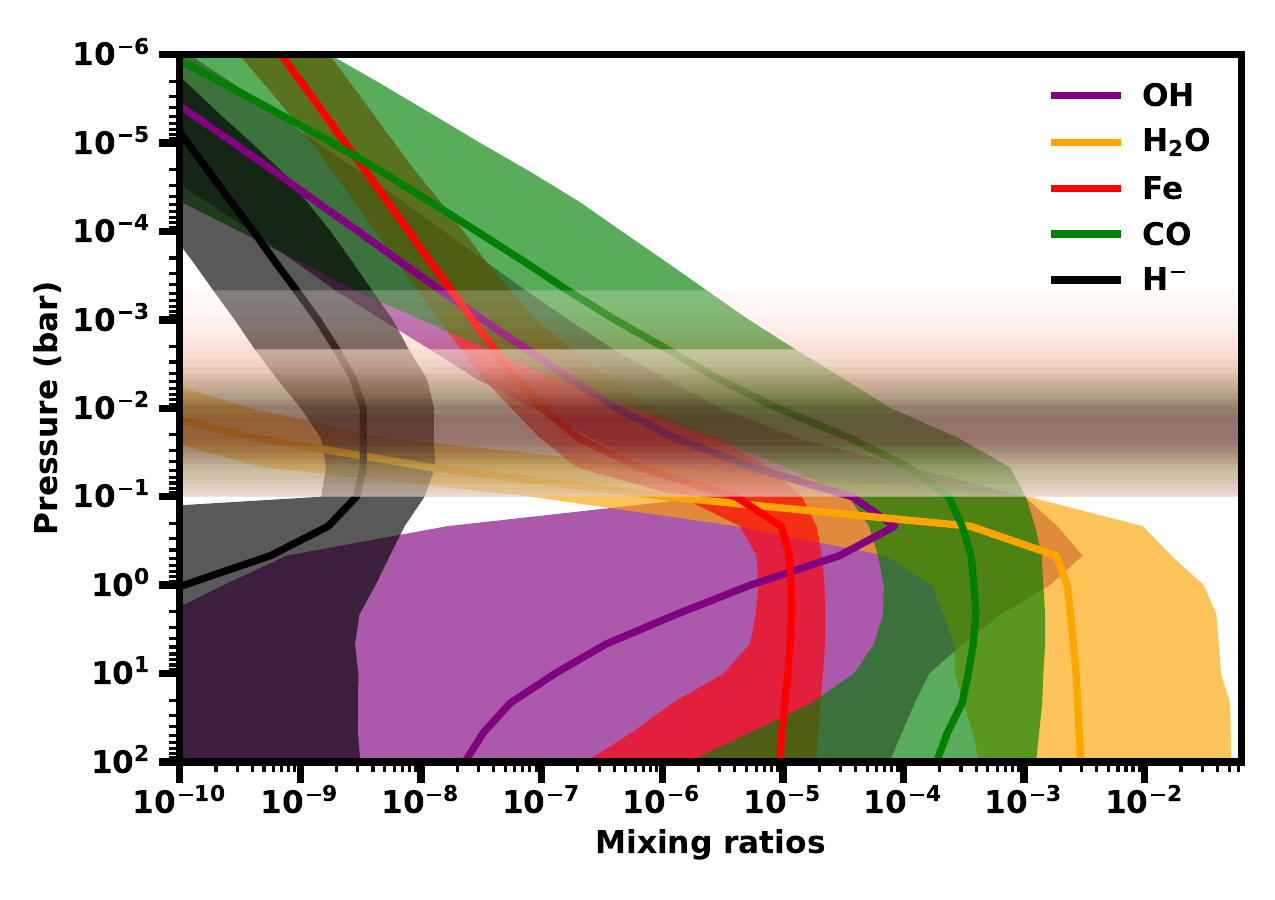} 
\caption{Derived vertical volume mixing ratio profiles for various species. These profiles are calculated using the retrieved $T$-$P$ profile together with the retrieved elemental abundance and metallicity. The solid line represents the median vertical volume mixing ratio profile, while the shadow shows the $1\sigma$ envelope. The horizontal shadows indicate the contribution functions of thermal emission from different species.
}
\label{Retri_OtoH_median_ABU}
\end{figure}

The corner plots of the retrieved elemental abundances are presented in Fig.~\ref{Retri_element_XH}. The [Fe/H] value of $-0.58_{-0.22}^{+0.26}$ tends towards subsolar to solar abundance, while the [O/H] value of $0.61_{-0.58}^{+1.19}$ implies solar to supersolar abundance. The [C/H] value of $0.07_{-0.67}^{+0.72}$ is broadly compatible with the solar value. The vertical volume mixing ratio profiles, shown in Fig.~\ref{Retri_OtoH_median_ABU}, are calculated using the retrieved $T$-$P$ profile, the elemental abundance, and [M/H] under the assumption of thermochemical equilibrium. In the pressure range corresponding to the infrared photosphere (i.e. $\sim$0.1 - 0.0005 bar, see Fig.~\ref{Retri_OtoH_median_ABU}), the volume mixing ratios of CO, H$_2$O, and Fe decrease rapidly with decreasing pressure, while the OH mixing ratio increases and then decreases rapidly due to thermal dissociation, resulting in a level comparable to those of CO, H$_2$O, and Fe. This suggests detectability for OH, CO, H$_2$O and Fe. However, despite its predicted high abundance, CO and H$_2$O was not clearly detected in our initial CCF analysis. This issue will be discussed in more detail in Section~\ref{RG}. 

\begin{figure}
\centering
\includegraphics[width=\hsize]{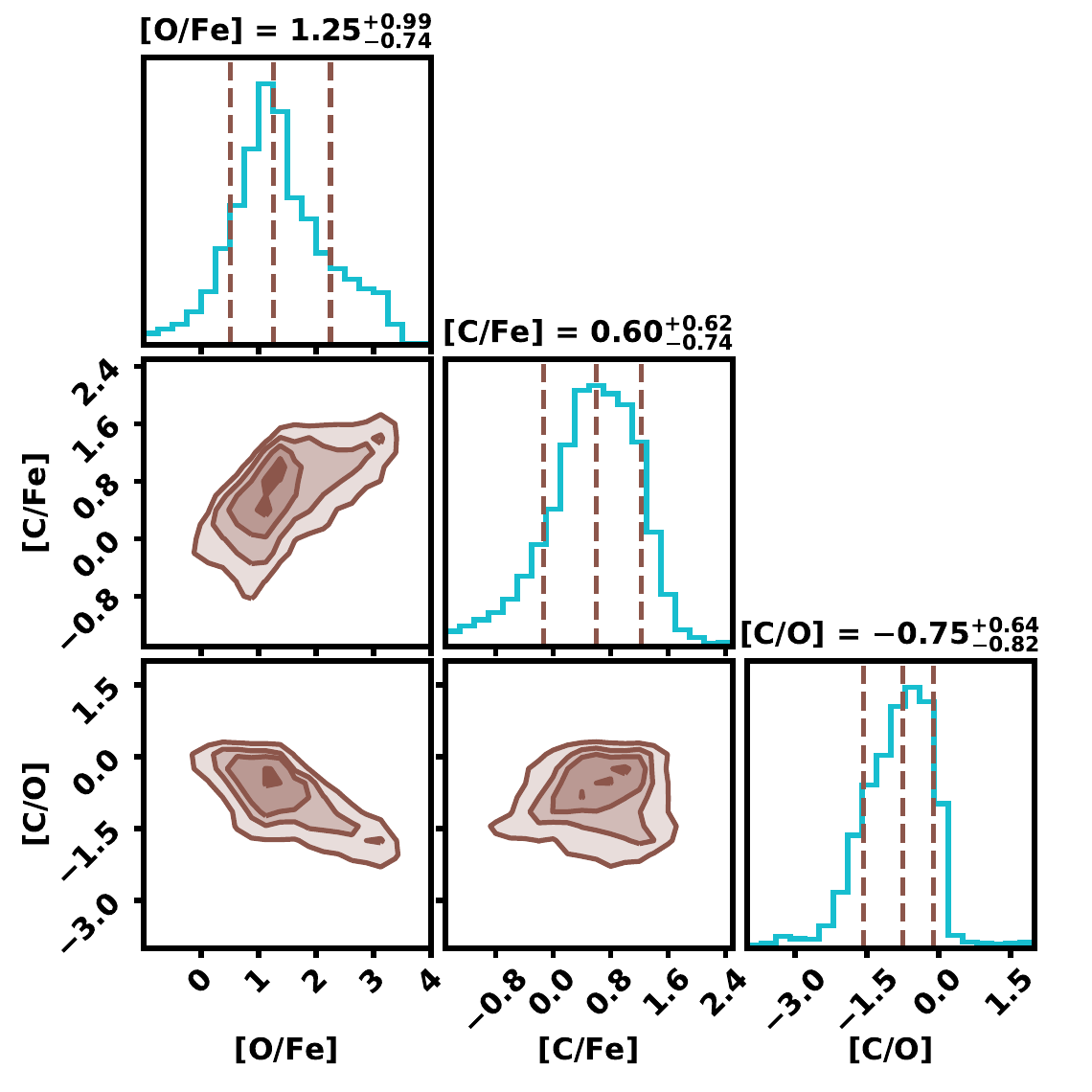}
\caption{
Same format as in Fig.~\ref{Retri_element_XH}, but showing the relative abundance of elements on a logarithmic scale.
}
\label{Retri_element_VR}
\end{figure}

The relative abundances of elements, derived from the retrieved elemental abundances, are shown in Fig.~\ref{Retri_element_VR}. The derived [C/O] of $-0.75_{-0.82}^{+0.64}$ falls within a subsolar to solar range (0.03$-$0.78 $\times$ solar). A solar to supersolar [O/H] value and a subsolar to solar [C/O] suggest that solid accretion, such as ice, occurred alongside gas accretion in the planet's envelope during its formation and migration \citep[e.g.][]{Mordasini+etal+2016,Madhusudhan+etal+2014,Madhusudhan+etal+2017,Cridland+etal+2019}. For classical hot Jupiter formation scenarios based on core accretion models, the planet likely formed beyond the water snowline and migrated inwards. Therefore, the solar to supersolar [O/H] value and subsolar to solar [C/O] suggest that the planet migrated through the protoplanetary disc while accreting planetesimals and/or drifting pebbles during the gas accretion phase or earlier, thereby retaining ice solids in its envelope. As the planet moved closer to its host star, the protoplanetary disc likely reached the sublimation temperature of certain species, primarily water ice, composed of volatile elements such as oxygen. This sublimation would have enriched the envelope's oxygen content, thereby lowering the [C/O] and increasing the [O/H] value \citep{Oberg+etal+2011,Mordasini+etal+2016,Kirk+etal+2024,Penzlin+etal+2024}. However, given the relatively weak constraints, alternative formation scenarios cannot be ruled out.

In addition, the volatile to refractory elemental ratio has emerged as a powerful tracer of giant‐planet formation and migration \citep{Lothringer+etal+2021}. Volatile species (e.g. C- and O-bearing molecules) condense beyond their snow lines, while refractory elements (e.g. Fe, Mg, Si) remain in solid form closer to the star. From our retrievals, we derive [O/Fe] and [C/Fe] values spanning solar to supersolar, suggesting possible volatile enrichment by accretion of ice-rich planetesimals. A high mass fraction of volatiles relative to refractory components (rock materials) implies that KELT-9 b accreted substantial icy solids in the outer disc before migrating inwards through a solid- and dust-poor region. This picture dovetails with our conclusions on planet formation and migration from [C/O] and [O/H], and the difference between [O/Fe] and [C/Fe] further shows the differential enrichment of oxygen-bearing versus carbon-bearing ices. Such volatile to refractory ratios derived with different elements can therefore constrain a planet's formation location relative to multiple snow lines and chart its subsequent migration pathway. Nevertheless, the current constraints are too weak to clearly distinguish between different formation scenarios, and these ratios can best provide tentative guidance on possible enrichment processes.

We retrieved [M/H] values of $-1.17_{-1.31}^{+1.64}$. Our results suggest that the metallicity of this planet is in the subsolar to solar range, which is broadly consistent with the results reported by \citet{Kasper+etal+2021} based on metal atoms and ions observed with Gemini-N/MAROON-X, as well as with the weakly constrained subsolar metallicity derived by \citet{Ridden-Harper+etal+2023} using CARMENES data. However, a notable discrepancy exists with the work of \citet{Jacobs+etal+2022}, who reported high atmospheric metallicity and subsolar [C/O] through self-consistent one-dimensional chemistry modelling. Their alternative model incorporating TiO/VO quenching yielded subsolar metallicity estimates similar to ours, albeit with different chemical assumptions. Different wavelength coverage and high- and low-resolution observations probe different pressure ranges, potentially explaining the discrepancies between our results and previous studies.

If the subsolar metallicity of this planet is correct, it might suggest a locally well-mixed envelope and core. This is because the metallicity of this planet is expected to correspond to a supersolar metallicity, as predicted by the metallicity-mass relation \citep{Thorngren+etal+2016,Thorngren+etal+2019,Mordasini+etal+2016}, while the observed low atmospheric metallicity implies the presence of compositional gradients in the envelope and interior \citep{Hasegawa+etal+2024,Swain+etal+2024}. However, metallicity alone does not fully describe the enrichment of individual elements within the planet, as it is scaled based on the overall ratios of metallic elements relative to the Sun or host star and does not account for differences in the enrichment or depletion of different elements (such as the oxygen enhancement and iron depletion we observed in KELT-9b, see Fig.~\ref{Retri_element_XH}). Processes such as ionization, cold traps, and escape fractionation can create such variations, especially in UHJs. Future studies should combine optical and infrared measurements to more thoroughly infer the abundances of various elements.

\section{Retrieval-guided species detection for high-resolution spectroscopy}\label{RG}

\begin{figure}
\centering
\includegraphics[width=\hsize]{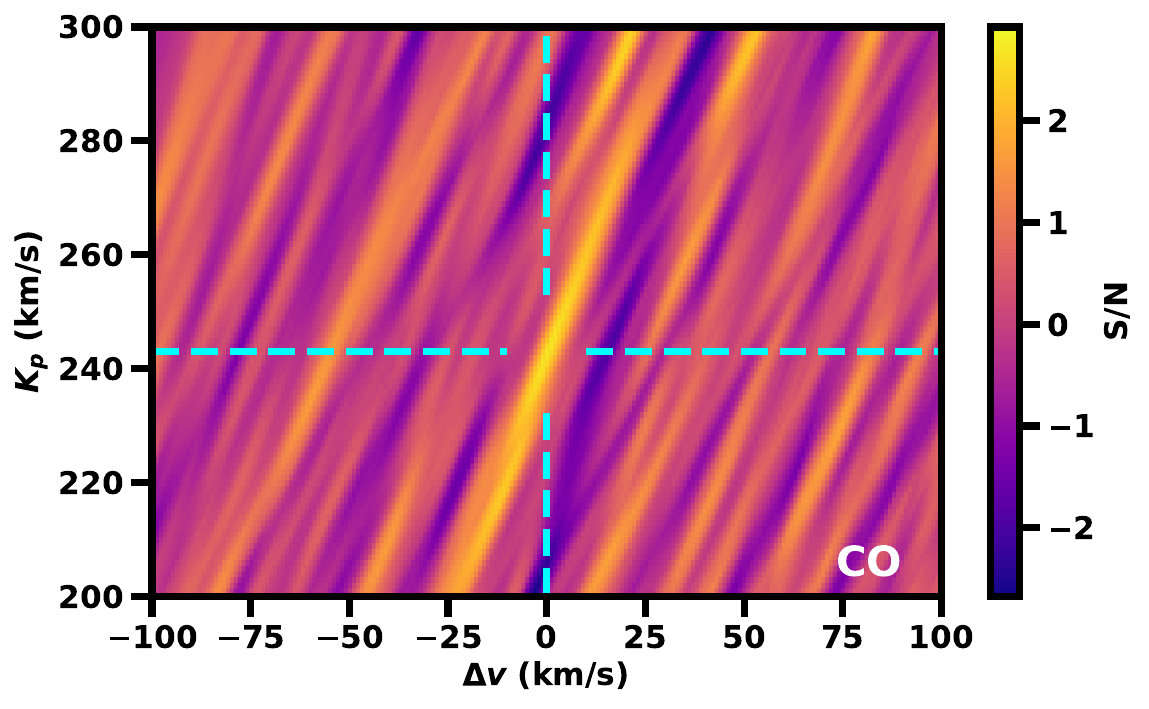} 
\caption{$K_\mathrm{p}$ map of CO based on the best-fitting model from the chemical equilibrium retrieval. The map is shown in the same format as the right panel of Fig.~\ref{CCF_FeOH}. The tentative detection demonstrates the utility of retrieval-guided cross-correlation analyses, highlighting subtle CO signals that were not detected in the initial analysis.
}
\label{kp_map}
\end{figure}

According to chemical equilibrium models, CO and H$_2$O are expected to be present in detectable abundances in the dayside atmosphere of KELT-9b. However, our initial CCF analysis, which utilized a custom-defined forward model, did not yield a clear detection of CO and H$_2$O. Motivated by this inconsistency, we adopted a less restrictive retrieval approach by expanding the set of chemical species included in the analysis, rather than limiting ourselves to only those species initially detected by CCF analysis. Specifically, our retrievals incorporated not only Fe and OH, which were robustly detected by CCF, but also additional species such as CO and H$_2$O (high-abundance species predicted by chemical equilibrium and strong cross sections in the instrument's wavelength range) that were not initially identified in the CCF results.

Interestingly, the retrieval analysis provided robust constraints on CO abundance, despite its non-detection in our initial CCF analysis. To further investigate this, we performed a dedicated CCF analysis using the best-fitting forward model obtained from the retrieval, restricting the chemical composition exclusively to CO. This approach led to the tentative identification of CO signals (see Fig.\ref{kp_map}) that were not detected in the initial CCF analysis. Nevertheless, the CCF signal of CO remains relatively weak, possibly due to the limited performance of SPIRou in the K-band, which leads to higher noise levels in the wavelength region where the ro-vibrational transitions of CO near 2.3~$\mu$m are located. For H$_2$O, the retrieval remains poorly constrained and even the best-fit model produces no significant CCF signal. This may indicate that H$_2$O has undergone strong thermal dissociation within the pressure range traced by our data, resulting in a low abundance (see Fig.~\ref{Retri_OtoH_median_ABU}).

In summary, we recommend a more inclusive retrieval-first strategy. Instead of restricting the choice of atmospheric species based solely on initial CCF detections, retrieval analyses should initially incorporate a wide variety of plausible chemical species, constrained only by physically justified parameters such as the planetary orbital velocity (i.e. $K_p$) and systemic radial velocity offset (i.e. $\Delta \varv$). Species that achieve strong posterior abundance constraints in the retrieval can then be independently verified through subsequent dedicated CCF analyses with the best-fitting forward model. This retrieval-guided cross-correlation strategy ensures a more unbiased and comprehensive utilization of high-resolution spectroscopic data, thereby enhancing the sensitivity and reliability of species detection in exoplanet atmospheres.

\section{Conclusions}\label{CONCL}

We conducted two observations of the dayside of KELT-9b using the SPIRou spectrograph. Through the application of the cross-correlation method, we report the detection of OH emission signals and confirm the presence of Fe in the planet's atmosphere. The simultaneous detection of both a molecule and an atom in this UHJ demonstrates SPIRou's exceptional capability for near-infrared molecular detection in exoplanetary atmospheres and highlights its potential as a valuable complement to the \emph{JWST}. Detection of OH, along with lack of a clear H$_2$O signature, aligns with the expected scenario of H$_2$O thermal dissociation on the dayside. Furthermore, we applied a retrieval framework to diagnose and constrain the atmospheric properties of this planet.

We employed a self-consistent chemical equilibrium atmospheric retrievals, constraining the elemental abundance and the metallicity. The retrieved $T$-$P$ profile revealed a significant thermal inversion layer in the dayside atmosphere, broadly consistent with previous optical observations, but located deeper in the atmosphere. This discrepancy may be due to the fact that infrared wavelengths probe deeper atmospheric layers than optical wavelengths. The rotational broadening is consistent with that expected from tidally locked rotation, and no significant blue- or red-shifted signals were detected to provide definitive insights into atmospheric dynamics. A detailed analysis of atmospheric dynamics requires consideration of the three-dimensional velocity field, the observed phases, and uncertainties in the systemic radial velocity. The retrieved subsolar to solar [C/O] ($-0.75_{-0.82}^{+0.64} \mathrm{dex}$) and solar to supersolar [O/H] ($0.61_{-0.58}^{+1.19} \mathrm{dex}$) suggest that this planet likely formed beyond the water snowline and migrated inward through the disc, accreting both solids and gas. Ratios of volatile to refractory elements spanning solar to supersolar likewise point out substantial accretion of ice-rich solids during the planet's formation and migration. However, given the relatively large uncertainties, these interpretations remain tentative, and alternative formation scenarios cannot be ruled out. The retrieval results also indicate that the subsolar to solar atmospheric metallicity could suggest a locally well-mixed envelope and interior. Further optical and infrared observations are necessary to measure the abundances of various elements and place more precise constraints on the formation and migration scenarios of UHJs.

Finally, although CO is expected to be abundant according to equilibrium chemistry, it was not detected in our initial CCF analysis. 
However, when the retrieval was performed with an expanded set of chemical species beyond those initially detected by CCF, CO was subsequently identified with strong posterior constraints. This result highlights that a strict ``CCF-first'' workflow may overlook marginal species and that a retrieval-guided cross-correlation strategy enables a more complete and unbiased atmospheric inventory.

\begin{acknowledgements}
      The authors thank the anonymous referee for the constructive comments and helpful suggestions on the manuscript.
      G.C. acknowledges the support by the National Natural Science Foundation of China (grant Nos. 42578016, 12122308, 42075122), 
      Youth Innovation Promotion Association CAS (2021315), and the Minor Planet Foundation of the Purple Mountain Observatory. 
      F.Y. acknowledges the support by the National Natural Science Foundation of China (grant no. 42375118). 
\end{acknowledgements}

\bibliographystyle{aa} 
\bibliography{manuscript.bib} 

\begin{appendix} 

\section{Additional figures}

\begin{figure*}[!b]
\centering
\includegraphics[width=\hsize]{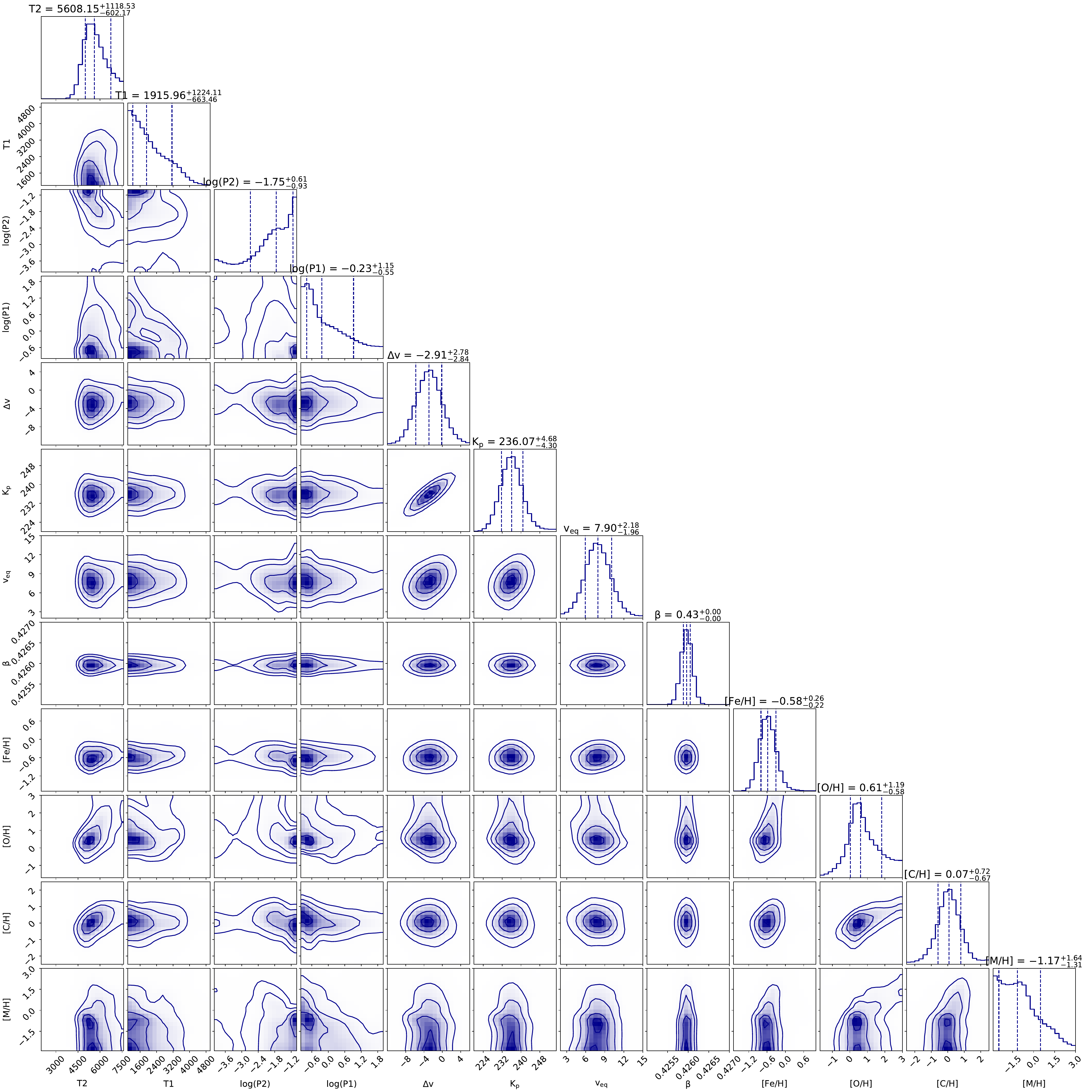}
\caption{Posterior distribution of the parameters from the MCMC fit combined two nights for chemical equilibrium retrieval.}
\label{Retri_OtoH_corner}
\end{figure*}

\begin{figure*}
\centering
\includegraphics[width=\hsize]{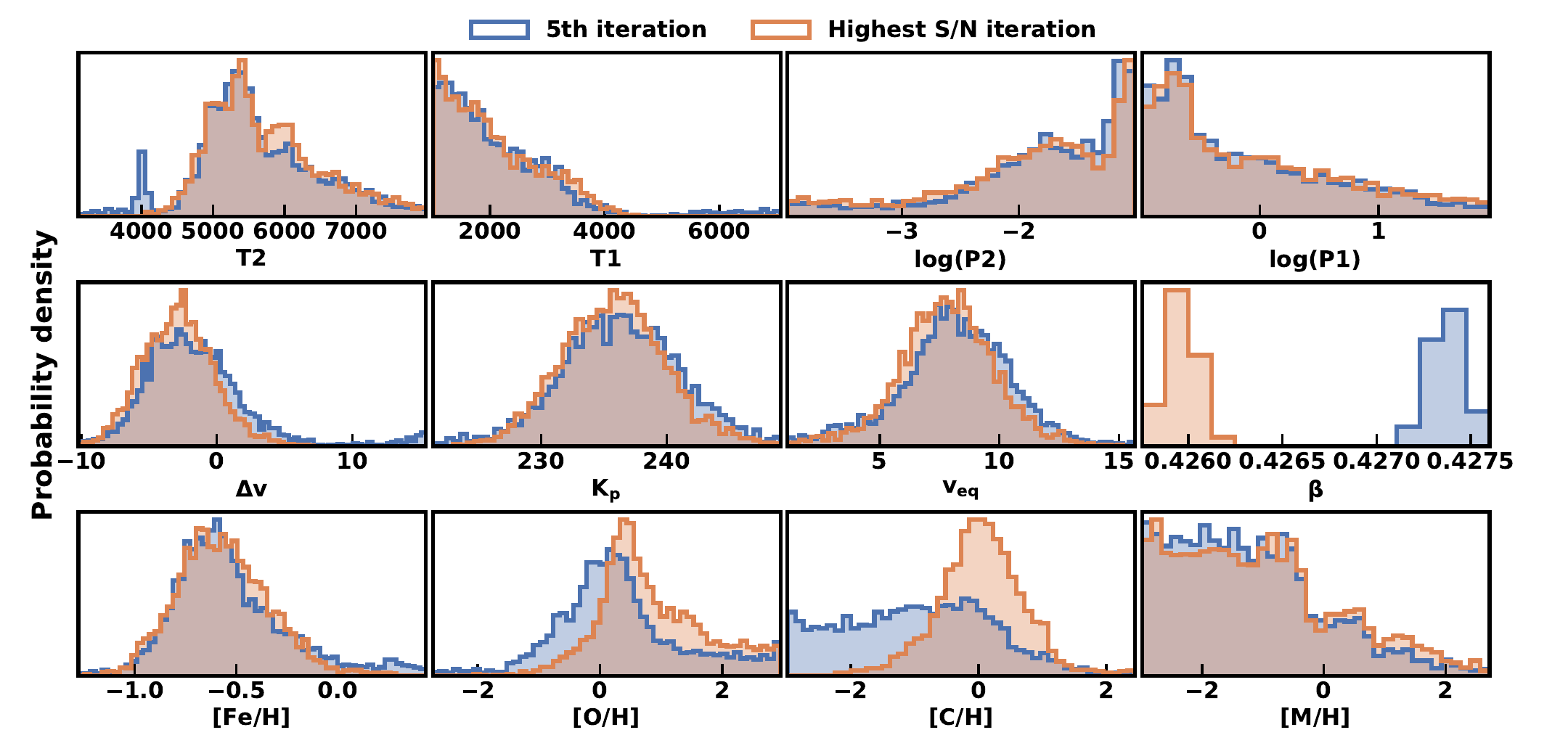}
\caption{Posterior distributions of the retrieved parameters from different SYSREM iterations. The blue corresponds to the 5th iteration, and the orange ones to the iteration with the highest S/N (i.e. the 4th and 10th iterations for Night-1 and Night-2, respectively). The parameters are consistent, with only the noise scaling factor $\beta$ showing slight variations due to SYSREM-based noise estimation.}
\label{Diff_iteration_marginals_compare}
\end{figure*}

\begin{figure}
\centering
\includegraphics[width=\hsize]{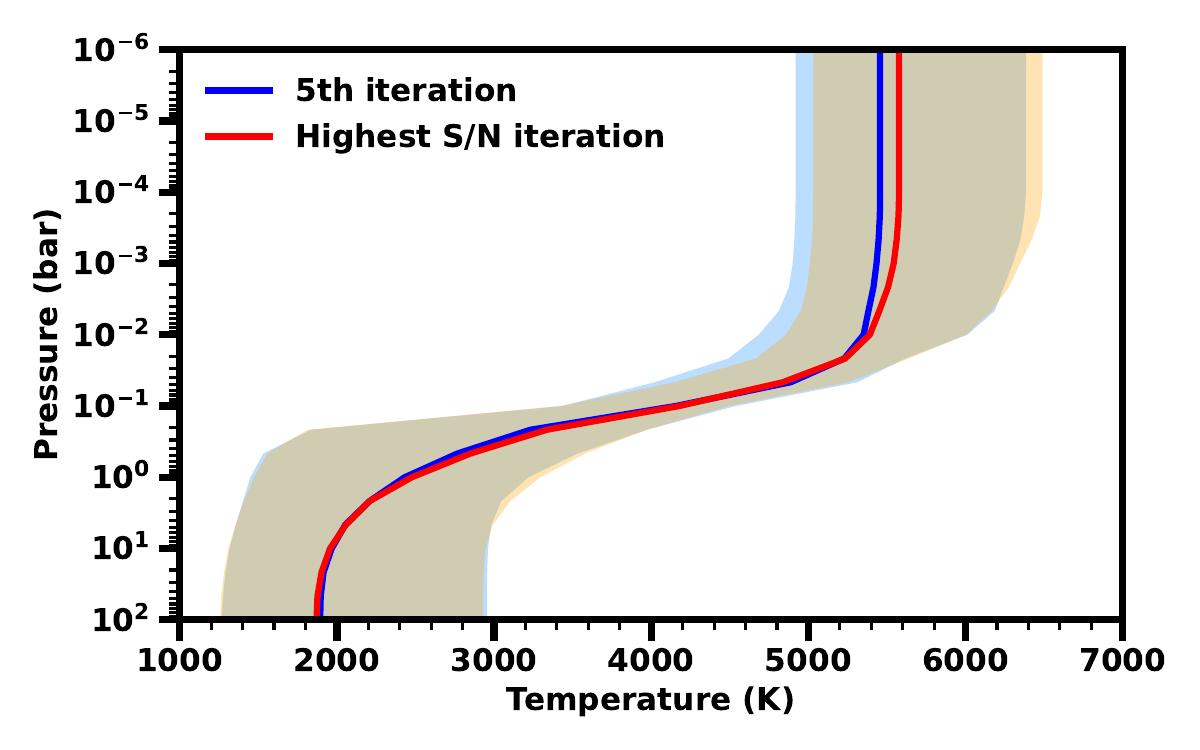}
\caption{Retrieved $T$-$P$ profiles from the chemical equilibrium retrieval using different SYSREM iterations. The blue line shows the 5th iteration and the red line corresponds to the iteration with the highest S/N (i.e. the 4th and 10th iterations for Night-1 and Night-2, respectively). The shaded regions indicate the 1$\sigma$ confidence intervals. The results indicate that the choice of SYSREM iteration has only a limited impact on our dataset.}
\label{Diff_iteration_TP_compare}
\end{figure}

\FloatBarrier
\section{Parameters of the KELT-9 system}

\begin{table}[htb!]
	\caption{Parameters of the KELT-9 system.}             
	\label{WASP76INFO}      
	\centering
	\resizebox{\linewidth}{!}{          
	\begin{tabular}{l c c}     
	\hline\hline 
	\noalign{\smallskip}      
		Parameter & Symbol [Unit] & Value \\
	\noalign{\smallskip}
	\hline\noalign{\smallskip} 
	\multicolumn{3}{l}{\emph{Stellar parameters}} \\
	\noalign{\smallskip}                
   		Stellar mass & $M_{\star}$ $[M_{\sun}]$ & $2.32 \pm 0.16$ $^{(a)}$ \\  
   		\noalign{\smallskip} 
		Stellar radius & $R_{\star}$ $[R_{\sun}]$ & $2.418 \pm 0.058$ $^{(a)}$ \\
   		\noalign{\smallskip} 
		Effective temperature & $T_\mathrm{eff}$ $[\mathrm{K}]$ & $9329 \pm 118$ $^{(a)}$ \\
		\noalign{\smallskip} 
		Systemic velocity & $\varv_\mathrm{sys}$ $[\mathrm{km}$ $\mathrm{s}^{-1}]$ & $-17.74 \pm 0.11$ $^{(b)}$ \\
	\noalign{\smallskip}
	\hline\noalign{\smallskip} 
	\multicolumn{3}{l}{\emph{Planetary parameters}} \\
		\noalign{\smallskip} 
		Planetary mass & $M_\mathrm{p}$ $[M_\mathrm{J}]$ & $2.88 \pm 0.35$ $^{(a)}$ \\ 		
		\noalign{\smallskip}
		Planetary radius & $R_\mathrm{p}$ $[R_\mathrm{J}]$ & $1.936 \pm 0.047$ $^{(a)}$ \\
		\noalign{\smallskip}  
        Planetary surface gravity & $g_{\mathrm{p}}$ $[\mathrm{m}~\mathrm{s}^{-2}]$ & $19.0_{-2.4}^{+2.5}$ $^{(a)}$ \\ 	
		\noalign{\smallskip}
		Equilibrium temperature & $T_\mathrm{eq}$ $[\mathrm{K}]$ & $3921_{-174}^{+182}$ $^{(a)}$ \\
		\noalign{\smallskip}
		Orbital semi-major axis & $a$ $[\mathrm{au}]$ & $0.03368 \pm 0.00078 $ $^{(a)}$ \\
		\noalign{\smallskip}
		Orbital period & $P$ $[\mathrm{d}]$ & $1.48111916 \pm 0.00000013$ $^{(c)}$ \\
		\noalign{\smallskip}
		Transit epoch (BJD) & $T_0$ $[\mathrm{d}]$ & $2458162.09128 \pm 0.00009$ $^{(c)}$ \\
		\noalign{\smallskip}
		RV semi-amplitude & $K_\mathrm{p}$ $[\mathrm{km}$ $\mathrm{s}^{-1}]$ & $234.24 \pm 0.90$ $^{(b)}$\\
		\noalign{\smallskip}
		Orbital inclination & $i$ $[\mathrm{deg}]$ & $86.79 \pm 0.25$ $^{(d)}$ \\
		\noalign{\smallskip}
	\noalign{\smallskip}
	\hline                  
	\end{tabular}}
	\tablebib{$^{(a)}$~\citet{Borsa+etal+2019}. $^{(b)}$~\citet{Hoeijmakers+etal+2019}. $^{(c)}$~\citet{Harre+etal+2023}.  $^{(d)}$~\citet{Gaudi+etal+2017}.}
	\end{table}

\end{appendix}

\end{document}